 \documentclass[preprint2]{aastex}

\usepackage{color}
\shorttitle{Implications of the Co-rotation Theorem
on the MRI in Axial Symmetry}
\shortauthors{Montani et Pugliese}

\newcommand{\tb}[1]{\textbf{#1}}

\newcommand{\il}{~}
\def\be{\begin{equation}}
\def\ee{\end{equation}}
\def\bea{\begin{eqnarray}}
\def\eea{\end{eqnarray}}

\begin{document}


\title{Implications of the Co-rotation Theorem
on the MRI in Axial Symmetry}


\author{G. Montani\altaffilmark{1}}
\affil{ENEA, FSN-FUSPHY-TSM, R.C. Frascati, Via E. Fermi 45, 00044 Frascati, Italy}
\email{giovanni.montani@frascati.enea.it}
\and
\author{F. Cianfrani}
\affil{Institute
for Theoretical Physics, University of Wroc\l{}aw, Pl.\ Maksa Borna
9, Pl--50-204 Wroc\l{}aw, Poland}
\and
\author{D. Pugliese}
\affil{Institute of Physics and Research Centre of Theoretical Physics and Astrophysics,  Faculty of Philosophy \& Science,
  Silesian University in Opava,
 Bezru\v{c}ovo n\'{a}m\v{e}st\'{i} 13, CZ-74601 Opava, Czech Republic}

\altaffiltext{1}{Physics Department, ``Sapienza'' University of Rome,\\{ P.le Aldo Moro 5, 00185 (Roma), Italy}}
%\altaffiltext{2}{Society of Fellows, Harvard University.}
%\altaffiltext{3}{present address: Center for Astrophysics,
%    60 Garden Street, Cambridge, MA 02138}
%\altaffiltext{4}{Visiting Programmer, Space Telescope Science Institute}
%\altaffiltext{5}{Patron, Alonso's Bar and Grill}

\begin{abstract}
We analyze the linear stability of an axially symmetric ideal
plasma disk, embedded in a magnetic field and endowed with a
differential rotation. This study is performed by adopting the
magnetic flux function as the fundamental dynamical
variable, in order to outline the role played by the
co-rotation theorem  on the linear
mode structure. Using some specific assumptions
(e.g. plasma incompressibility and propagation
of the perturbations along the background
magnetic field), we select the Alfvenic nature of
the Magneto-Rotational Instability
and, in the geometric optics limit,
we determine the dispersion relation
describing the linear spectrum.
We show how the implementation of the co-rotation
theorem (valid for the background configuration) on the linear dynamics produces the
cancellation of the vertical derivative of
the disk angular velocity
(we check such a feature also in the standard
vector formalism to facilitate comparison with
previous literature, both in the axisymmetric and three-dimensional case). As a result, we
clarify that the unstable modes have, for a
stratified disk, the same morphology, proper of a
thin disk profile, and the $z$-dependence
has a simple parametric role. 
\end{abstract}

\keywords{accretion, accretion disks; plasmas; magnetohydrodynamics (MHD)}
\section{Introduction}
In the study of stellar accretion disks \\\citep{B01},
as well as in the general problem of the accretion
profile around compact astrophysical sources
\citep{Abr11}, the question concerning the stability of the rotating
plasma is central in the understanding of the
angular momentum transport.
The rotation of the accretion
system is ensured by the gravitational field
of the astrophysical source and it is not too far
from the Keplerian value, at least for thin disk
configurations \citep{Ol97}.
Another basic aspect of the accreting
structures consists of their axial symmetry, which can
reduce their equilibrium and the associated
linear perturbation scheme to a pure two-dimensional problem.
However,
a rotating fluid, having a nearly Keplerian differential
rotation profile, is clearly stable under perturbations which
preserve its axial symmetry \citep{Ray17,BH98}. This issue,
still recently debated \citep{PAlA,Balbnature}, is of crucial relevance
in the set up of the original Shakura idea
for accretion \citep{S73,SS73}, based on the
angular momentum transport across the disk by means of
shear viscosity, associated with the differential
rotation.
In fact, the requested value of the
viscosity coefficient, to account for the
observed accretion rates, especially in highly compact
stars, like X-ray Pulsars, is not justified by the kinetic
predictions in the typical range of density and temperature
available for such sources.
The instability of the plasma profile is then necessary to
account for the onset and the full establishment of
a turbulent regime; it can be   easily restated as a laminar
motion in the presence of an effective viscosity, generated
via the non-zero correlation functions of the turbulent
fluid velocity field components, see \citep{BH2003}  and also \citep{Montani:2012wr,
Montani:2012xt}.
As well known, the solution to the puzzle of a stable
rotating quasi-Keplerian plasma has been provided by including
the presence of a weak magnetic field, always present in
astrophysical accreting systems, which is able to
trigger within the axial picture, a continuum of
unstable Alfvenic modes, known as the Magneto-rotational
instability (MRI) \citep{V59,C60}.
It can be shown that MRI is
able to generate a turbulent flow
and a satisfactory effective shear viscosity amount \citep{BH91,BH98,BH2003}. A discussion of the MRI in the case of a
stratified axially symmetric plasma disk can
be found in  \citep{Ba:1995} (and also \citet{BU}),
 where
it is shown how the unstable modes are triggered
by the gradient profile of the angular velocity,
differently from the non-magnetized configuration,
for which the stability criterion relies on the
specific angular momentum \cite{Ray17}.
In this study perturbations of the steady state are considered adiabatic, accordingly to the local and Boussinesq
approximation. The system of equations for the linear
evolution is closed by using the entropy equation,
in which the perturbed pressure can be negligible, as in
a typical problem of the so-called internal waves, when
the restoring force, acting on the small disturbances, is
mainly due to the gravitational field of the central object.

In the present analysis, we address the
same problem, but focusing our attention
on the specific Alfvenic character of the
MRI and, as the main feature of the
considered approach, we investigate the implication that the validity of the
co-rotation theorem \citep{F37}
(the background disk angular frequency must be a function of the magnetic flux
function) has on the linear mode spectrum (see also \citep{MP13}).
In particular, we consider a real incompressible plasma in the disk
(hypothesis corresponding to a polytropic index approaching infinity in the
equation of state) and we restrict
the study to the  perturbations propagating along the magnetic field only.
We first face the analysis of the perturbed linear dynamics, by using a
formalism in which the magnetic flux function is a basic dynamical variable
and
the dispersion relation is calculated in
the geometrical optic limit
(at all equivalent to the local
approach adopted in \citep{Ba:1995}).
Then, in order to shed light on the basic
novelty, introduced by the co-rotation theorem, we repeat the same
linear perturbation scheme
in the usual scenario, relying on the
use of the magnetic vector components as
in \citep{Ba:1995}. We stress how, in this
paper, we consider a pure poloidal
background magnetic field, but we then discuss how the role of the azimuthal background magnetic field does not change this picture.

In such vector formulation, the co-rotation condition stands as the
orthogonality relation between the background magnetic field and the
corresponding gradient of the disk angular
velocity.

The main issue of our study consists in
demonstrating how the implementation of
the co-rotation condition, on the linear perturbation dynamics, leads to the
cancellation of the vertical derivative
of the angular frequency from the dispersion relation. As a direct consequence
the
linear mode spectrum retains the same form
 for a thin disk configuration
(where the $z$-dependence does not enter
the disk profile). This result implies that the vertical profile of the disk
affects only in a parametric way the
stability properties of the steady axisymmetric configuration.
Thus, we are lead to attribute more
relevance to other instability mechanism, see for instance \citep{Coppi(2008)} and \citep{BU},
when the vertical gradients are so marked
to affect the disk steady configuration. Finally, in order to deepen the coupling between the constant poloidal background magnetic
field and the perturbation profile we consider non-axisymmetric disturbances, according to the vector formalism. We show how the stability
condition is affected by the toroidal wave number and MRI results suppressed when the latter is sufficiently large compared to the modulus of the poloidal wave vector.

More specifically in Sec.\il(\ref{Sec:fis-iss}), we provide the basic formalism at the ground of
our analysis, by specifying all the dynamical equations characterizing the
ideal MHD scenario. In Sec.\il(\ref{Sec:lpt}), we define
the general properties of the steady background configuration and then we
construct the linear perturbation scheme,
determining the evolution of small disturbances affecting the background
profile.
In Sec.\il(\ref{Sec:GOL}), we address the geometric
optic limit to extract from the linear
dynamics the dispersion relation governing
the mode spectrum and we discuss the
obtained issue.
In Sec.\il(\ref{Sec:VF}), the calculation of the
normal modes is repeated in the standard
scheme, relying on the use of the poloidal magnetic field component, instead
of the magnetic flux function, here
considered. In Sec.\il(\ref{Sec:nnas}), we analyze the non-axisymmetric case in the presence of a constant toroidal background magnetic field
component, as allowed by the co-rotation profile. The dispersion relation is derived according to the vector formulation of the previous section
and its implications are commented.
Concluding remarks follow.
\section{Basic Equations}\label{Sec:fis-iss}
Let us consider the system
made of the Faraday law and of the electron force balance
\begin{eqnarray}
\label{flefb}
\partial_t \vec{B}=- c\vec{\nabla} \wedge \vec{E}\\
\vec{E} + \frac{\vec{v}}{c}\wedge \vec{B}=0
\label{flefb2}
\, ,
\end{eqnarray}
where $\vec{E}$ and $\vec{B}$ denote the
electric and the magnetic fields respectively.
Equation
(\ref{flefb}) admits the solution
\begin{equation}
\vec{E}=- \vec{\nabla} \Phi - \frac{1}{c}
\partial_t \vec{A}
\, ,
\label{potvec}
\end{equation}
where $\vec{A}$ is the vector potential,
such that the magnetic field $\vec{B}=\vec{\nabla}\wedge \vec{A}$,
while $\Phi$ denotes the time dependent electric potential
with
$\vec{\nabla}\cdot \vec{A}=0$ (Coulomb gauge).
Using this solution in (\ref{flefb2}), we obtain
\begin{equation}
\vec{\nabla} \Phi + \frac{1}{c}\partial_t \vec{A}
=\frac{\vec{v}}{c}\wedge \vec{B}
\, .
\label{eqf}
\end{equation}
In what follows, we consider a two-dimensional
axisymmetric system, labeled by cylindrical
coordinates $\{ r,\phi ,z\}$; all the
functions involved are independent
of the azimuthal angle $\phi$.
Without loss of generality, we now
express the magnetic field in the form
\begin{equation}
\vec{B}=-\frac{1}{r}\partial_z\psi \vec{e}_r +
\frac{\bar{B}_{\phi}}{r}\vec{e}_{\phi} +
\frac{1}{r}\partial_r\psi \vec{e}_z
\, ,
\label{mgf}
\end{equation}
or equivalently, we take the vector potential
\begin{equation}
\vec{A}=\vec{A}_p +
\frac{\psi }{r}\vec{e}_{\phi}
\, ,
\label{potv}
\end{equation}
where, in the Coulomb gauge, the
poloidal component $\vec{A}_p$ satisfies
the two conditions
\begin{equation}
\vec{\nabla} \wedge \vec{A}_p=\frac{\bar{B}_{\phi}}{r}\vec{e}_{\phi}
\quad ; \quad
\vec{\nabla} \cdot \vec{A}_p=0
\, .
\label{potvp}
\end{equation}
We also consider the following general velocity field
in the plasma
\begin{equation}
\vec{v}=\vec{v}_p + \omega r\vec{e}_{\phi}
\, ,
\label{vlf}
\end{equation}
where $\vec{v}_p=v_r\vec{e}_r + v_z\vec{e}_z$
denotes the poloidal velocity field.
Separating the azimuthal and the poloidal
components of the Eq.\il(\ref{eqf})
we obtain
\begin{eqnarray}
\partial_t \psi + \vec{v}_p \cdot
\vec{\nabla} \psi=0
\label{bea} \\
\vec{\nabla} \Phi + \frac{1}{c} \partial_t
\vec{A}_p=\frac{\omega}{c} \vec{\nabla} \psi +
\frac{1}{c}\vec{v}_p\wedge (\vec{\nabla} \wedge \vec{A}_p)
\label{beb}
\, .
\end{eqnarray}
We rewrite this system as two
scalar equations, by taking the curl of
Eq.\il(\ref{beb}) (having the azimuthal component only), together with (\ref{bea})
 we consider the  equation
\begin{eqnarray}
\nonumber
\partial_t \bar{B}_{\phi} + \vec{v}_p \cdot
\vec{\nabla} \bar{B}_{\phi}  + \bar{B}_{\phi}\vec{\nabla} \cdot
\vec{v}_p-2\frac{\bar{B}_{\phi} v_r}{r}=r(\partial_z\omega \partial_r \psi
- \partial_r\omega \partial_z\psi ).
\\\label{beb1}
\end{eqnarray}
It is worth stressing that Eq.\il(\ref{bea}) is
a gauge invariant relation, being the azimuthal
component of the (physical) electron force
balance equation.
The azimuthal component of the ideal
MHD momentum conservation provides the
following equation for the angular velocity
\bea\nonumber
\rho r\left(\partial_t \omega + \vec{v}_p\cdot
\vec{\nabla} \omega \right)
+2\rho v_r\omega=\\
\frac{1}{4\pi r^2}
\left(\partial_r\psi \partial_z\bar{B}_{\phi}
- \partial_z\psi \partial_r\bar{B}_{\phi} \right)
\, ,
\label{bei}
\eea
where $\rho$ is the mass density of the plasma profile.
The poloidal component of the momentum conservation
equation takes the form
\begin{eqnarray}
\label{poeq}
\rho \left(\partial_t\vec{v}_p +
\vec{v}_p\cdot \vec{\nabla}\vec{v}_p
- \omega^2r\vec{e}_r\right)=- \vec{\nabla}p -
\nonumber \\
-\frac{1}{4\pi r}\left[ \partial_r\left(\frac{1}{r} \partial_r\psi\right)
+ \frac{1}{r}\partial^2_z\psi
\right] \vec{\nabla}\psi -
\nonumber \\
- \frac{1}{8\pi r^2}\vec{\nabla}\bar{B}_{\phi}^2 + \vec{F}^e_p
\, ,
\end{eqnarray}
where $p$ denotes the pressure and by $\vec{F}^e_p$
we indicate the external forces acting on the
plasma configuration, which is retained non-zero
in the meridian plane only.
We complete this scheme by the continuity
equation, describing the mass conservation
via the behavior of the mass density $\rho$
\begin{equation}
\partial_t\rho + \vec{\nabla}\cdot
\left(\rho \vec{v}_p\right)=0
.
\label{coeq}
\end{equation}
The structure of this system of equations,
characterizing the axisymmetric plasma evolution,
(\ref{bea})-(\ref{bei}),
allows to fix some important points.
i)-From Eq.\il(\ref{beb1}), it is immediate
to recognize that if the azimuthal magnetic field
 vanishes,
then we must have $\omega=\omega (\psi )$.
%However,  this hypothesis
%on the form of $\bar{B}_{\phi}$
%can be significantly relaxed,
%as soon as we take into account Eq. (\ref{bea}).
%In fact, when we require $\bar{B}_{\phi}=\bar{B}_{\phi}(\psi )$
%and the plasma incompressibility, the
%left-hand-side of Eq. (\ref{beb1})
%identically vanishes, and it implies $\omega=\omega(\psi)$.
%This is the non-stationary extension of the Ferraro's
%co-rotation theorem \citet{F37}.
%Clearly, when $\bar{B}_{\phi}$ identically vanishes,
%the  co-rotation theorem holds  for a
%compressible fluid too.
ii)-Eqs.\il(\ref{beb1}) and (\ref{bei}) show
how the angular velocity $\omega$ and the azimuthal
magnetic field $\bar{B}_{\phi}$ have a strict correlation,
since they are able to generate  each other.
In fact, if one of these two quantities is
constant or a function of  $\psi$, the other one has
a vanishing right-hand-side in its dynamical
equation:  we then get
 linear homogeneous equations in normal form,
having the null solution as the unique one,
in correspondence to a vanishing initial value.
Thus, if the right-hand-side of Eqs.\il(\ref{beb1}) and (\ref{bei})
vanishes, the variable $\bar{B}_{\phi}$ and $\omega$ respectively
can not be generated from the dynamics if they vanish
in the beginning.

\section{Linear perturbation theory}\label{Sec:lpt}
We now develop a linear perturbation
approach around a background
configuration, corresponding to a
purely rotating  plasma disk, embedded
in a vacuum poloidal magnetic field\footnote{The magnetic
field of a compact star is typically
well-modeled by a dipolar configuration \citet{PM92}.}.
Here we denote by a suffix $(0)$ all the background
quantities and by the suffix $(1)$ all the
corresponding linear fluctuations.
The main  assumption we adopt in our
analysis is the  Alfvenic nature of the
perturbation
i.e.
 $\vec{\nabla}\cdot \vec{v}_{p1}=0$
(an Alfvenic mode does not transport matter).
As effect of the plasma backreaction,
the magnetic {surface function} admits the
following natural decomposition
$\psi=\psi_0 + \psi_1$, with
$\mid \psi_1\mid \ll \mid \psi_0\mid$.
In the case of a purely rotating background
configuration, Eq.  (\ref{bei}) is
automatically verified at the zero order,
while the poloidal system (\ref{poeq})
naturally splits in two background equations\footnote{Indeed the
external force coincides  with the star gravity, i.e.
\(
\vec{F}_0^e=-\omega_K^2(r,z^2)\vec{r}_p\), \(\vec{r}_p=(r,z)
\),
$\omega_K^2=GM_s/(r^2 + z^2)^{3/2}$,
being the Keplerian angular frequency
(here $G$ denotes the Newton's constant and $M_s$
the mass of the central body).}
\begin{eqnarray}
\label{greq}
\vec{\nabla}p_0=\rho_0 \left(\omega_0^2(\psi_0)r\vec{e}_r -
\omega_K^2(r,z^2)\vec{r}_p\right) \\
\label{pore}
\frac{1}{4\pi r}\left[\partial_r\left(\frac{1}{r} \partial_r\psi_0\right)
+ \frac{1}{r}\partial^2_z\psi_0
\right]=0
\, .
\end{eqnarray}
The first of these equations is the gravostatic
equilibrium determining the disk morphology,
while the second one is the force-free condition for
the vacuum magnetic field of the central object.
Since for the static axisymmetric background
the co-rotation theorem holds, we take
the angular velocity in the form
$\omega=\bar{\omega}(\psi) +
\omega^+$, i.e. we separate a contribution
depending on the function $\psi$ at any order
from a generic angular velocity term.
Clearly, we have $\omega^+_0=0$ and
\begin{equation}
\omega_1=\frac{d\omega}{d\psi}
\mid_{\psi=\psi_0}\psi_1 +
\omega^+_1 \equiv \bar{\omega}_1+ \omega^+_1, \; \bar{\omega}_1
\equiv\dot{\omega}_0(\psi_0)\psi_1.
\label{omeg}
\end{equation}
In order to address the perturbation scheme, we
introduce the poloidal plasma shift $\vec{\xi}_p$,
defined via the relation
$\vec{v}_{1p} \equiv \partial_t \vec{\xi}_p$.
Perturbing Eq. (\ref{bea}), we
get the basic relation
\begin{equation}
\partial_t\psi_1 + \vec{v}_{1p}\cdot
\vec{\nabla}\psi_0=0 \, \rightarrow
\psi_1=-\vec{\xi}\cdot \vec{\nabla}\psi_0
\, .
\label{eeqq}
\end{equation}
Eq. (\ref{omeg}) allows,
together with (\ref{eeqq}),
to write the perturbed azimuthal momentum conservation (\ref{bei})
in the  form
\begin{equation}\label{psxi}
r\partial_t\omega^+_1=-2\omega_0v_{1pr} +
\frac{1}{4\pi r^2\rho_0}\left(\partial_z(\bar{B}_{\phi})_1\partial_r\psi_0 - \partial_r(\bar{B}_{\phi})_1
\partial_z\psi_0 \right).
\end{equation}
It is worth stressing that Eq. (\ref{eeqq})
permits to cancel out the contribution of
$\bar{\omega}_1$, simply because its variation
is induced by the perturbation $\psi_1$
(actually, we used above the relation
$\vec{\nabla}\omega_0=\dot{\omega}_0
\vec{\nabla}\psi_0$).
The first order structure of Eq. (\ref{beb1})
takes the simple form
(the contribution due to $\bar{\omega}_1$  cancels out)
\begin{equation}
\partial_t(\bar{B}_{\phi})_1=r\left(\partial_z\omega^+_1
\partial_r\psi_0 - \partial_r\omega^+_1\partial_z\psi_0
\right)
\, .
\label{eebb}
\end{equation}
Observing that in the linear perturbation regime,
the induced poloidal magnetic field remains much smaller
than the background component, i.e.
$\mid \vec{\nabla}\psi_1\mid \ll \mid \vec{\nabla}\psi_0\mid$,
the radial and vertical equations take, at the first
order, the form
\begin{eqnarray}
\nonumber
&&\rho_0 \left[
\partial_t^2\xi_r - 2\omega_0 r
\left(\dot{\omega}_0\psi_1 + \omega^+_1\right) \right]=
\\
&&\nonumber=
\frac{\rho_1}{\rho_0}\partial_rp_0-\partial_rp_1 -
\frac{1}{4\pi r}\Delta \psi_1
\partial_r\psi_0 \\
\nonumber
&&\rho_0 \partial^2_t\xi_z
=\frac{\rho_1}{\rho_0}\partial_zp_0- \partial_zp_1 -
\frac{1}{4\pi r}\Delta \psi_1
\partial_z\psi_0,
\\
\label{beq1}
\end{eqnarray}
where we introduced the notation
\begin{equation}\label{lapl}
\Delta \psi_1 \equiv \partial_r\left(\frac{1 }{r}\partial_r\psi_1 \right) +
\frac{1}{r}
\partial^2_z\psi_1
\, .
\end{equation}
Finally, since we requested the incompressibility of the plasma,
from the perturbed mass conservation Eq. (\ref{coeq}), by time integration, we have
\begin{equation}\label{repr}
\rho_1=- \vec{\xi}_p\cdot \vec{\nabla} \rho_0
\, .
\end{equation}
The behavior of the perturbed pressure
$p_1$ will be deduced via the
preservation of the incompressibility
along the evolution (see below).
Taking the second time derivative of $\psi_1$
from Eq. (\ref{eeqq}), using Eqs.
(\ref{beq1}) to express the
corresponding second time derivative of the shift
vector components, we get the relation
\begin{eqnarray}\label{qfe}
\partial^2_t\psi_1 + 2\omega_0r\partial_r\omega_0
\psi_1 + 2\omega_0r\partial_r\psi_0 \omega^+_1=\\
\frac{1}{\rho_0}\vec{\nabla}\psi_0\cdot
\vec{\nabla}p_1
+ v_A^2 r\Delta \psi_1
\nonumber
\, ,
\end{eqnarray}
where $v_A^2=v_{Ar}^2 + v_{Az}^2$ is the
square of the background Alfven speed, with
\begin{equation}\label{alve}
v_{Ar}^2=\frac{\left(\partial_z\psi_0\right)^2}
{4\pi r^2\rho_0},\quad
v_{Az}^2=\frac{\left(\partial_r\psi_0\right)^2}
{4\pi r^2\rho_0}
\, .
\end{equation}
Eq. (\ref{qfe}), together with
 (\ref{eeqq}), (\ref{psxi}) and (\ref{repr}),
constitute the system of perturbed equations,
able to provide the dispersion relation
for the corresponding spectrum of modes.
\section{Geometric optic limit}\label{Sec:GOL}
In what follows, we shall address the
geometric optic limit for the perturbed quantities
$(...)_1$, which are taken in the form
\begin{equation}
\label{pefo}
(...)_1(t,r,z)=(...)^+(t,r,z) \exp \{ i \Theta (t,r,z) \}
\, ,
\end{equation}
where $(...)^+$ is a small and regular
(smooth like the background)
amplitude,
while the function $\Theta$ is a very large phase
(since it varies of $2\pi$ on the small
perturbation wavelength).
Therefore, we will naturally introduce
the definitions
\(
\vec{k} \equiv \vec{\nabla}\Theta
\), \(\Omega \equiv -\partial_t\Theta
%\label{keom}
\)
so that, we get the basic relation of the
adopted approximation, i.e.
\begin{equation}\label{pdko}
\partial_t(...)_1=-i\Omega (...)_1
\, , \quad
\vec{\nabla} (...)_1=i\vec{k}(...)_1
\, .
\end{equation}
This approach to the perturbation analysis is equivalent to a local approximation
(large wavenumber limit), but the
request of a linear theory, for which
the perturbed magnetic field must be small in comparison to the background one, implies the restriction

\begin{equation}
\mid \vec{k}\mid |\psi_1| \ll
\mid \vec{\nabla}\psi _0\mid
\, .
\label{rest}
\end{equation}
Indeed, we retained
$\vec{\nabla}\psi _0$ in Eq.
(\ref{eeqq}) (otherwise no evolving
perturbations would arise), as well as
the gradient $\vec{\nabla}\rho _0$
in Eq. (\ref{repr}).
Using such approximation scheme,
i.e. retaining dominant terms
in the wavenumber vector $\vec{k}$
(as well as the inhomogeneous term in the
angular velocity, responsible for MRI), Eqs.\il
(\ref{beq1}) take
the compact form
\begin{equation}
\Omega ^2 \vec{\xi}_p +
2\omega _0\left( \dot{\omega}_0\psi _1 +
\omega ^+_1\right) \vec{e}_r
= i\vec{k}\frac{p_1}{\rho _0} - \frac{k^2\psi _1}{4\pi
r^2\rho _0} \vec{\nabla}\psi _0
\, .
\label{beqk}
\end{equation}
In order to select the Alfenic character of the MRI, we now choose the wavevector
$\vec{k}$ along the background field $\vec{B}_0$, so eliminating
the magnetic pressure contribution.
In our formalism, such a condition reads as $\vec{k}\cdot \vec{\nabla}\psi _0 = 0$
and this simplifies the expression of the
perturbed pressure $p_1$, as calculated from Eq. (\ref{beqk}). In fact, taking
the scalar product by $\vec{k}$ and implementing the incompressibility condition
$\vec{k}\cdot \vec{\xi}_p = 0$, we get
\begin{equation}
k^2p_1
	=-
2i\rho_0\omega _0k_rr\left(
\dot{\omega}_0 \psi _1 + \omega ^+_1\right)
\, .
\label{rho1}
\end{equation}
This relation states the preservation of
the plasma incompressibility in the linear evolution of the perturbations.\footnote{ If we had evaluated this same condition from Eqs. (\ref{beq1}), the gradient $\vec{\nabla}\rho _0$ would have added, via Eq. (\ref{repr}), the term
$\Omega ^2\rho _1$ in the left-hand-side of Eq. (\ref{beqk}). Such a contribution is negligible for an incompressible fluid because of the large value of the sound velocity $c_s^2 \equiv \gamma p_0/\rho _0$, with the polytropic index $\gamma \rightarrow \infty$.}
Analogously,
Eq. (\ref{qfe})
rewrites
\begin{equation}\label{qfef}
\left(\Omega^2 - y_r - \omega_A^2\right)\psi _1
=2\omega_0r\partial_r\psi_0 \omega^+_1
\, ,
\end{equation}
where $y_r \equiv 2\omega_0r\partial_r\omega_0$,
$\omega_A^2\equiv k^2 v_A^2$.
Eqs. (\ref{psxi}) and (\ref{eebb})
read as
\begin{eqnarray}
\label{psx1}
r\Omega \omega^+_1=-2\omega_0\Omega \xi_r
- \frac{\vec{k}\cdot \vec{B}_0}{4\pi \rho_0r}(\bar{B}_{\phi})_1 \\
\label{eeb1}
\Omega (\bar{B}_{\phi})_1=- r^2\vec{k}\cdot \vec{B}_0\omega^+_1
\, .
\end{eqnarray}
Combining together these last two equations,
we easily get
\begin{equation}
r\left(\Omega^2 - \bar{\omega}_A^2\right)\omega^+_1
=- 2\omega_0\Omega^2\xi_r
\, ,
\label{omxi}
\end{equation}
Finally, the first of  (\ref{beq1}) provides the relation
\begin{equation}
\Omega^2\partial_r\psi_0\xi_r=-\left(\alpha y_r + v_{Az}^2k^2\right) \psi_1
-2\alpha \omega_0r\partial_r\psi_0 \omega^+_1
\, ,
\label{xiom}
\end{equation}
where $\alpha \equiv 1 - k_r^2/k^2$.
Using Eq. (\ref{xiom}) into
the relation (\ref{omxi}), we can express the
quantity $\omega^+_1$ in terms of $\psi_1$ and
then Eq. (\ref{qfef}) yields the dispersion relation
\bea \label{dire}
&&\Omega ^4-{ b}\;\Omega^2
+ { c}  = 0,
\\\nonumber
&&{{b}}\equiv \left( K_0^2 + 2\omega _A^2
\right) + 4\omega _0^2 (\alpha-1)\quad{ c}\equiv \omega _A^2\left( y_r + \omega _A^2\right),
\eea
where $K_0^2\equiv y_r +4\omega _0^2$ is
the epicyclic frequency and $\omega ^2_{Ar}
\equiv \omega _A^2 - k^2v_{Az}^2<\omega _A^2$.
It can be proved that
at the
necessary condition to get MRI is
provided by the condition
$\omega _A^2 < - y_r$, which ensures
 $c < 0$  (in fact the position $b < 0$ requires again $c < 0$).

 \section{Vector formulation}\label{Sec:VF}

In order to better elucidate how the validity of the
co-rotation theorem for the background configuration influences
the structure of the dispersion relation, we here  analyze
 the linear perturbation dynamics, using the
same vector formulation adopted in \citet{Ba:1995}.

We start by writing down the basic evolution equations for
the linear corrections in terms of the poloidal
magnetic field $\vec{B}_p$ and the toroidal component
$B_{\phi}$, as well as by using the perturbed poloidal
velocity field $\vec{v}_{1p}$ (absent in the background), instead of the plasma shift.
The equations below are determined under the
same hypotheses of the previous sections and we also
introduce the perturbed toroidal velocity ${v_{\phi}}_1$,
without any additional splitting between its co-rotational
and generic parts.

The incompressibility condition clearly stands as
$\vec{k}\cdot \vec{v}_{1p} = 0$, while the poloidal
components of the momentum conservation take the form

\bea&&\nonumber
\Omega \vec{v}_{1p} - 2i\omega _0
v_{\phi_1 }
\vec{e}_r
-\frac{\vec{k}}{\rho _0}p_1 -i\frac{\vec{\nabla} p_0}
{\rho _0^2}\rho _1
\\
&&
-\frac{1}{4\pi \rho _0}\left(
\vec{B}_0\cdot \vec{B}_1\vec{k}
- \vec{k}\cdot \vec{B}_0\vec{B}_1\right) = 0
\, ,
\label{va}
\eea

where $\vec{B}_1$ denotes the perturbed poloidal
component of the magnetic field.

The poloidal component of the induction
equation correspondingly gives

\begin{equation}
\Omega \vec{B}_1 = -
\vec{k}\cdot \vec{B}_0\vec{v}_{1p}
\, .
\label{va1}
\end{equation}

Substituting this relation in Eq.
(\ref{va}) and using the incompressibility constraint
(since $\vec{k}$ is parallel to
$\vec{B}_0$, then $\vec{B}_0\cdot
\vec{v}_{1p} = 0$), we rewrite it as follows

\begin{equation}
\left( \Omega ^2 - \omega _A^2\right)
\vec{v}_{1p}
= 2i\Omega \omega _0 \vec{e_r}v_{\phi_1} +
\frac{\Omega \vec{k}}{\rho _0}p_1.
\label{va2}
\end{equation}

Above, we neglected the term containing
$\rho _1$, coherently with the mass conservation equation. Furthermore,
preserving the incompressibility
condition leads again to determine the
perturbed pressure $p_1$ as in Eq.(\ref{rho1}), which substituted back yields
\begin{equation}
\left( \Omega ^2 - \omega _A^2\right)
\vec{v}_{1p}
= 2i\Omega \omega _0v_{\phi_1}
\left(\vec{e}_r - \frac{k_r}{k^2}\vec{k}
\right).
\label{va2x}
\end{equation}

Now, the azimuthal components of
the momentum conservation equation
and the induction one, respectively
write
\begin{equation}
\Omega v_{\phi_1} + ir
\vec{v}_{1p}\cdot \vec{\nabla}
\omega _0 + 2i\omega _0v_{1pr}
+ \frac{\vec{k}\cdot \vec{B}_0}
{4\pi \rho _0}B_{\phi 1} = 0
\,
\label{va3}
\end{equation}

and

\begin{equation}
\Omega B_{\phi 1} -
i {r}\vec{B}_1\cdot \vec{\nabla}\omega _0
+ \vec{k}\cdot \vec{B}_0 v_{\phi_1} = 0
\,.
\label{va4}
\end{equation}

Substituting in this last Eq.
(\ref{va4}) the expression of $\vec{B}_1$
in terms of $\vec{v}_{1p}$, as
provided by Eq. (\ref{va1}), we
can easily get the
$B_{\phi 1}$, to be inserted into
Eq. (\ref{va3}). So doing, we arrive
to the following basic relation

\begin{equation}
\left( \Omega ^2 - \omega _A^2\right)
\left[ i\Omega v_{\phi_1} - r\vec{v}_{1p}
	\cdot \vec{\nabla}\omega _0
\right]	= 2\Omega^2\omega _0 v_{1pr}
	\, .
	\label{va5}
\end{equation}

We now multiply Eq. (\ref{va2x})
by the vector $\vec{\nabla}\omega _0$
and recalling that, under the considered
hypotheses, it is $\vec{k}\cdot \vec{\nabla}
\omega _0 = 0$, we get the relation

\begin{equation}
\left(\Omega ^2 - \omega _A^2 \right)
r\vec{v}_{1p}\cdot \vec{\nabla}\omega _0
= i\Omega y_rv_{\phi_1}
\,.
\label{va6}
\end{equation}

The equation above can be substituting
in Eq. (\ref{va5}), yielding

\begin{equation}
\left(\Omega ^2 - \omega _A^2 - y_r\right)v_{\phi} = - 2i\Omega \omega _0v_{1pr}
\, .
\label{va7}
\end{equation}

Finally, the radial component
of Eq. (\ref{va2x}) reads as

\begin{equation}
\left( \Omega ^2 - \omega _A^2\right)
v_{1pr} = 2i\Omega \omega _0\alpha v_{\phi_1}
\, .
\label{va8}
\end{equation}

Combining together Eqs. (\ref{va7})
and (\ref{va8}) we easily recover the
dispersion relation of Eq.\il(\ref{dire}).

Let us now discuss how the present picture can be affected by
the presence of a background toroidal magnetic field.
Actually, including such a background magnetic field
component, in the perturbation scheme, would
correspond to modify Eq. (\ref{va}) only
(see also \citet{Ba:1995}, where this component
is included ab initio), by adding a term of
the form $-B_{\phi 0}B_{\phi 1}\vec{k}/4\pi \rho _0$.
This contribution is clearly the additional toroidal
pressure component, the tension modification vanishes
because of the axial symmetry prescription, removing
the azimuthal component of the wavevector.
However, when we calculate the perturbed pressure from
the preservation of the incompressibility along the
plasma evolution and then substituting it back into
Eq. (\ref{va}), the additional term naturally cancell out of
the poloidal momentum conservation equations.
Since that stage, the background poloidal magnetic field
component disappears from the perturbation scheme
and hence from the dispersion relation
(clearly in such a case, the wavevector is parallel
to the poloidal magnetic field only). 

\section{Non-axisymmetric perturbations}\label{Sec:nnas}
In this section, we face the study of
non-axisymmetric Alfvenic perturbations,
including in the background configuration a non-zero toroidal component of the
magnetic field $B_{\phi 0}$, along the
scheme traced in section
\ref{Sec:GOL}.
In \cite{Balbus92}, the same problem
is analyzed in the case $\omega _0 =
\omega _0(r)$, outlining how the dynamo
mechanism implies a time dependence of
$B_{\phi 0}$ and, hence, a time
dependence of the perturbation wave-numbers too (see also \cite{GL}).
However, when the co-rotation condition is preserved for the background
({\it i.e.} we deal with a stationary, axisymmetric and purely rotating configuration),
Eq. (\ref{eebb}) clearly admits the only
solution $B_{\phi 0} = const.$.
Thus, as an important consequence of the
co-rotation profile, no dynamo effect
takes place and we deal with a stationary
background even for a three-dimensional
(non-axisymmetric) problem.

Thus, we now consider perturbations, whose plane wave representation
is characterized by the term
$\exp \{ i(\vec{k}\cdot \vec{r}_p + m\phi )\}$, where
$m$ is an integer number. Then, introducing the notation

\begin{eqnarray}
\vec{\mathcal{B}}_0 \equiv \left( \vec{B}_0\, ,\, B_{\phi 0}
\right)
\quad , \quad
\vec{\kappa} \equiv \left( \vec{k}\, ,\, \frac{m}{r}
\right)
\nonumber \\
\bar{\omega}_A^2 \equiv \frac{\left(\vec{\kappa}
\cdot \vec{\mathcal
{B}}_0\right)^2}{4\pi \rho_0}
\, , \, \Omega^* \equiv \Omega + m\omega _0
\, .
\label{sa}
\end{eqnarray}

Eqs (\ref{va2}) and (\ref{va5}), once restated for the perturbed velocity
vector $\vec{v} = (\vec{v}_p\, ,\, v_{\phi 1})$,
can be expressed in the following single vector equation
(we note that, in the geometric optics limit, we have
$m\gg1$)

\begin{eqnarray}
i\Omega^*\left( \Omega^{*2} - \bar{\omega}_A^2 \right) \vec{v}_1 =
i\frac{\Omega^{*2}\vec{\kappa}}{
\rho _0}\left(p_1 + p_1^{(m)}\right) -
\nonumber \\
- 2\Omega ^{*2}\omega _0\left( \vec{e}_r
	v_{\phi 1} - \vec{e}_{\phi}v_{pr1} \right) +
\nonumber \\
+\left(\Omega ^{*2} - \bar{\omega}_A^2\right)
r\vec{v}_p\cdot \vec{\nabla}\omega _0
\vec{e}_{\phi}
\,,
\label{sb}
\end{eqnarray}

where $p^{(m)}_1 \equiv
\vec{\mathcal{B}}_0\cdot \vec{\mathcal{B}} 1/4\pi$ denotes the perturbed magnetic
pressure, $\vec{\mathcal{B}}_1
= (\vec{B}_{p1}\, ,\, B_{\phi 1})$
being the full perturbed magnetic field.

Since, we are considering an incompressible fluid,
{\it i.e.} $\vec{v}_1\cdot \vec{\kappa} = 0$, the pressure
$p_1$ can be calculated by preserving such a constraint in
the momentum conservation equation above and, then substituted there, so getting

\begin{eqnarray}
i\Omega^*\left( \Omega^{*2} - \bar{\omega}_A^2 \right) \vec{v}_1 =
-2\Omega ^{*2}\omega _0\left[
\left( \vec{e}_r - \frac{k_r}{\kappa ^2}\vec{\kappa}\right) v_{\phi 1}\right] +
\nonumber \\
+ \left[\left(\Omega ^{*2} - \bar{\omega}_A^2\right)
r\vec{v}_p\cdot \vec{\nabla}\omega _0
+ 2\Omega ^{*2}\omega _0v_{p1r}\right]
\left(\vec{e}_{\phi}
- \frac{m}{r\kappa^2}\vec{\kappa}\right)
\,,
\label{sc}
\end{eqnarray}

where $\kappa^2 \equiv k^2 +
m^2/r^2$.

Since, we still consider $\vec{\kappa}$ parallel to
$\vec{\mathcal{B}}_0$, then we have
$\vec{\kappa}\cdot \vec{\nabla} \omega _0
= \vec{k}\cdot \vec{\nabla}\omega _0 = 0$.
Hence, Eq. (\ref{va6}) rewrites as

\begin{equation}
\left( \Omega ^{*2} - \bar{\omega}_A^2\right)
r\vec{v}_p\cdot \vec{\nabla}\omega _0 =
i\Omega ^*y_r v_{\phi }
\, .
\label{sc2}
\end{equation}

Substituting this relation in the azimuthal component
of Eq. (\ref{sc}), we easily reach the following
generalization of Eq. (\ref{va7})

\begin{equation}
\left( \Omega^{*2} - \bar{\omega}_A^2 - \alpha _{\phi} y_r
+2i \Omega^* \omega_0\beta _{r\phi}\right) v_{\phi 1} = -2i\Omega ^*\omega _0\alpha _{\phi}
v_{pr1}
\, ,
\label{sc3}
\end{equation}

where $\alpha _{\phi}\equiv 1 - m^2/r^2\kappa ^2$ and $\beta _{r\phi}\equiv m k_r/r \kappa^2$.

Finally, the radial component of Eq. (\ref{sc})
gives the complementary relation

\begin{eqnarray}
\left( \Omega^{*2} - \bar{\omega}_A^2 -
2i\Omega ^*\omega _0\beta _{r\phi} \right) v_{pr1} =
\nonumber \\
 = \left( 2i\Omega ^*\omega _0\alpha - \beta _{r\phi}y_r\right)
v_{\phi 1}
\, .
\label{sc4}
\end{eqnarray}

Now, the
dispersion relation takes the form

\begin{eqnarray}
&&{\Omega^*}^4-{ \tilde{b}}\;{\Omega^*}^2
+ { \tilde{c}}  = 0,
\\\nonumber
&&{{\tilde{b}}}\equiv 2\bar{\omega}_A^2 + \alpha _{\phi}
	y_r + 4
\omega _0^2\frac{k_z^2}{\kappa^2},
\\\nonumber
&& {{\tilde{c}}}\equiv \omega _A^2\left(\alpha_{\phi} y_r + \omega _A^2\right)
\label{fff}
\end{eqnarray}

The study of the stability is qualitatively the same as for (\ref{dire}),  which can be easily recovered for $m\equiv 0$. In particular,
the necessary condition to get MRI is now $\omega _A^2 < - \alpha_{\phi} y_r$, which ensures
 $c < 0$ ($b < 0$ requires again $c < 0$). Therefore, it is easy to recognize that the non-axisymmetric MRI is suppressed when
\begin{equation}
\frac{m}{r} \gg |\vec{k}_p|,
\end{equation}
while in the opposite case we recover the stability condition of the previous sections (axisymmetric case).

\section{Discussion and Conclusions}
We analyzed the morphology that the MRI
takes in two-dimensional axial symmetry,
when the validity of the co-rotation theorem for the background configuration
of a stratified differentially rotating disk is
taken into account. We studied
an incompressible plasma, subjected
to the further restrictions
(non affecting the Alfvenic nature of the
MRI) that the background azimuthal magnetic field vanishes identically and
the perturbations propagate
along the background poloidal magnetic
field. The dispersion relation for the
normal modes  is derived in the geometrical optic limit and it turns out to
be isomorphic to that one of a thin
disk configuration.

By other words, also in the case of a
stratified disk the relevant quantity in
triggering the MRI is the radial gradient
of the disk angular velocity, while the
vertical profile of the background configuration does not enter the unstable
mode spectrum, apart from a parametric
dependence of all the quantities involved
in the problem.

It is worth noting that, with respect to
the analysis developed in \cite{Ba:1995},
we do not address the Boussinesq approximation and do not use the entropy
evolution
equation. Nonetheless, the behavior of
the perturbed mass density is negligible
in both the approaches and therefore
they clearly overlap (indeed, for an incompressible
plasma, the polytropic index approaches
infinity and the entropy equation for the perturbations essentially reduces to the mass conservation
law). The reason for
a net discrepancy in the two studies of
the stratified disk stability, according to
the ideal MHD representation,
is in our accounting for the co-rotation theorem. Indeed, when the wave
vector is parallel to the background poloidal magnetic field, the momentum
conservation equations must
preserve the spatial constraint
$\vec{k}\cdot \vec{\nabla}\psi _0 = 0$
(in the vector form it reads
$\vec{B}_0\cdot \vec{\nabla}\omega _0=0)$,
whose existence affects the dispersion
relation.
Clearly, the morphology of the background
has a direct impact on the nature and the
propagation of the linear disturbances.
The direct dependence of the background
angular velocity on the magnetic flux surfaces represents a well-defined link
between the magnetic configuration and
the differential rotation of the stratified disk, having the key implication
to cancel the vertical derivative of
the disk angular frequency in the dispersion relation.

Finally, we observe, how in the limit of a thin disk
configuration, the co-rotation theorem does not affect the
MRI profile and, actually, we obtain the well-know standard
result (see\cite{BH98},
\cite{BH91}). This is because the condition
$\vec{B}_0\cdot \vec{\nabla}\omega _0=0$ is identically
satisfied: for a thin disk we have $B_{r0}\sim 0$
(the profile for a compact star is typically dipole-like)
and $\omega _0 = \omega _0(r)$ (the vertical dependence
can be properly averaged out).

For the sake of completeness in Sec.\ref{Sec:nnas} we discuss also non-axisymmetric perturbations, compatible with a
co-rotation profile. In fact, as far as $\omega=\omega(\psi)$ only a constant poloidal magnetic field is permitted. We show how
the dispersion relation implies MRI suppression as soon as the poloidal wave vector components is sufficiently large, {\it i.e.} much
greater than the poloidal component. This fact has intriguing implications on the disk morphology since the validity of such a condition
depends on the ratio $m/r$ and therefore it is better fulfilled in the inner disk regions.

Although, the present analysis aims to outline the
implications of the co-rotation theorem on the MRI
morphology, according to a stationary and axisymmetric
background (corresponding to the same assumptions
at the ground of \citet{Ba:1995}), it is worth addressing the
question concerning if a \tb{ real} disk fulfills such conditions.
The problem is, in principle complicated, depending on the specific accretion process, i.e. accretion of the compact
object from a surrounding nebula, a companion binary star, etc.,
but a simple theoretical consideration can shed light on the
reliability of the presented scenario.
The first point to be focused is
that the validity of the present analysis
requires that the considered assumptions,
i.e. the predictivity of the co-rotation
theorem, are valid for a time scale
much greater than the growth time of the
MRI, namely $\omega_0^{-1}$.

Actually, the validity of the co-rotation theorem,
apart from the axial symmetry (implicit in the gravitationally confined disk configuration)
requires that the system is stationary and the poloidal
velocity vanishes, see Eq. (\ref{beb1}) \citet{Benini:2011dd}.

Let us discuss these two prescriptions 
if the radial profile is concerned, which indeed is the only 
relevant contribution for the accretion and, as we have seen above, 
for the MRI mode spectrum.

To deal with the co-rotation theorem, the radial
in-falling velocity must be much smaller than the
azimuthal one, i.e. $\mid v_r\mid \ll \omega _0r$.
Such a condition can be refined, by
using Eq. (\ref{beb1}), which yields
\begin{equation}
\mid v_r\mid \ll
\frac{r}{H}\frac{B_p}{B_{\phi}}
\omega_0 r
\, ,
\label{est}
\end{equation}
where $H$ is the disk half-depth,
while $B_p$ and $B_{\phi}$ denote
the estimate of the poloidal and azimuthal
background field intensity, respectively.
As far as $B_{\phi}$ does not
exceed $B_p$ (we recall that the magnetic
field of the central object is dipole-like), we see that
$\mid v_r\mid \ll \omega_0 r$ is a more
constraining condition.

Furthermore, the stationarity request implies that the
radial fluid acceleration be much smaller than the
gravitational one, namely
$\mid v_r\mid /\tau _{ns} \ll \omega _K^2 r$, where
$\tau_{ns}$ denotes the typical time scale for the emergence
of a system non-stationarity.

The first of these requirements is ensured by the proportionality
relation between the radial velocity and the viscosity
coefficient of the plasma, i.e. $\mid v_r\mid \propto
\eta_v$.

However, before MRI develops, the turbulence in
the disk cannot be triggered by powerful linear instabilities
and the viscosity coefficient $\eta_v$ has essentially
the very small kinetic value, ensured by the ion-ion
collisions\citep{Montani:2012xt}.
Thus, the assumption of a pure rotating background disk
is conceptually well-grounded.

The second requirement must be evaluated in correspondence to
the shortest time scale that we can postulate for the non-stationarity,
i.e. the MRI growth time, $\tau_{ns}\approx \tau _{MRI}\sim \omega _0^{-1}$.
Hence we get: $\mid v_r\mid \ll \omega _K^2 r/\omega _0 =
\omega _Kr(\omega _K/\omega _0)$. By other words,
as far as the disk angular velocity is not much greater
than the Keplerian one, the first requirement implies the
second too. Actually, in a stratified disk, these two
angular velocities can differ from each other, but their
ratio cannot violate the proposed scheme, due to
the smallness of $\mid v_r\mid$, i.e. due to the weakness
of the initial plasma disk viscosity
(for a discussion of the allowed values
for the above ratio
in a thick disk, see \cite{Ol97}).

It is worth stressing that, the radial velocity remains
small in comparison to the azimuthal one even in the
viscous $\alpha$-Shakura disk
\citet{S73,B01}, in
which context, the MRI must be implemented to account for the continuous
trigger of the
turbulence.
In particular, we note that in a
typical viscous time $\tau _v\sim r/\mid v_r\mid$ (the time in which the
accreting material falls from a given $r$
to the central object), the MRI
fully develops, since $\omega _0\tau _v\gg1$, as a consequence of the first
requirement.

Then the validity of the
co-rotation theorem, when studying the MRI, is a well-grounded
assumption,  but when the accretion process is strongly
enhanced (for instance in the collapsar profile
 or in the cataclysmic variables),
in which case the stationarity hypothesis can be also
clearly questioned.

\acknowledgments

This work has been developed in the framework of the CGW Collaboration
(www.cgwcollaboration.it). DP wishes to thank the Blanceflor Boncompagni-Ludovisi, n\'ee Bildt Foundation, for support during the first  development of this work,  and thanks  the Accademia
Nazionale dei Lincei (within the Royal Society fellowship program) for support during 2015, and acknowledges support from the Junior GACR grant of the Czech Science Foundation No:16-03564Y. FC is supported by funds provided by the National Science Center under the agreement
DEC-2011/02/A/ST2/00294.

\end{document}